# Data-Driven Prognosis of Failure Detection and Prediction of Lithium-ion Batteries

Hamed Sadegh Kouhestani[1], Lin Liu[1,*], Ruimin Wang[1], and Abhijit Chandra[2]

[1]University of Kansas, Department of Mechanical Engineering, 3136 Learned Hall, 1530 W. 15th St., Lawrence, KS 66045-4709, United States of America

[2]Iowa State University, Department of Mechanical Engineering, 2106 Black Engr, 2529 Union Dr., Ames, IA 50011-2030, United States of America

(*Corresponding Author: linliu@ku.edu, Ph: +1-785-864-1612, Fax: +1-785-864-5254)

**Abstract**

Battery prognostics and health management predictive models are essential components of safety and reliability protocols in battery management system frameworks. Overall, developing a robust and efficient battery model that aligns with the current literature is a useful step in ensuring the safety of battery function. For this purpose, a multi-physics, multi-scale deterministic data-driven prognosis (DDP) is proposed that only relies on in situ measurements of data and estimates the failure based on the curvature information extracted from the system. Unlike traditional applications that require explicit expression of conservation principle, the proposed method devices a local conservation functional in the neighborhood of each data point which is represented as the minimization of curvature in the system. By eliminating the need for offline training, the method can predict the onset of instability for a variety of systems over a prediction horizon. The prediction horizon to prognosticate the instability, alternatively, is considered as the remaining useful life (RUL) metric. The framework is then employed to analyze the health status of Li-ion batteries. Based on the results, it has demonstrated that the DDP technique can accurately predict the onset of failure of Li-ion batteries.

## 1.0 Introduction

Li-ion batteries (LIBs) are becoming ubiquitous in the energy storage units for plug-in or full electric vehicles (EVs). Based on the statistics obtained by Electric Drive Transportation Association (EDTA), EV sales in the United States market have increased from 345 vehicles in 2010 to 601,600 in 2022, with a total of 1.8 million EVs over the twelve-year sales period [1].

This trend has also been observed globally as EV sales reached 2.1 million in 2019 worldwide, boosting the stock to 7.2 million [2]. Although the outbreak of the COVID-19 pandemic dramatically affected the global EV market, prompting the market to plummet over the year 2020 relative to 2019, the projections exhibit an increase in EV stock from 7.2 million to 14 million in 2025 and 25 million in 2030, which accounts for 10% of global passenger vehicle sales in 2025 and 28% in 2030 [2, 3]. This rapid growth is attributed to the LIBs superior characteristics over lead-acid, nickel-cadmium, and nickel-metal-hydride cells, such as lower weight, higher energy density, relatively low self-discharge rate, and longer life cycle [4]. These features and their low emissions impact have played a significant role in the vast adoption of LIBs in various applications, especially in the transportation sector [5].

Although EVs' market is witnessing an unprecedented evolution, fast adoption of these vehicles requires more thorough status analysis of the battery performance's functionality and reliability as the primary power source and energy storage unit for EVs. Owing to their rechargeable nature, LIBs operation is subject to different irreversible processes taking place during their charging and discharging cycles and causing capacity fade due to various degradation mechanisms such as formation of solid-electrolyte interphase (SEI) on the surface of the electrodes, lithium decrease in electrodes caused by lithium plating or electrolyte oxidation which negatively impacts the cell performance [6-10]. Other deteriorating factors can also impact the LIBs degradation, such as electrode decay originating from variations of the volume of active materials during cycles, which will result in induced mechanical stress, a decrease in lithium density of storage sites, and chemical decomposition of electrodes [11]. These cycles severely deteriorate the battery's electrochemical and mechanical constituents, leading to power loss and capacity fade [12, 13]. In general, these processes result in battery capacity degradation, which usually results in battery failure, with consequences ranging from operational loss, downtime, and catastrophic malfunctions [14].

To address the aforementioned issues, over the recent years, numerous studies have been dedicated to proposing proper degradation model mechanisms for improving the reliability and availability of LIBs [15]. However, current degradation models suffer from inadequate detailing of the capacity models, mostly due to the high complexity and computational burden associated with nonlinear models [16]. Furthermore, in some cases, the effect of current and state of charge on the degradation models may be overlooked, resulting from data insufficiency to develop a detailed degradation model and the complexity of the computational models [17]. On the other hand, failure of LIBs does not necessarily occur due to a single battery degradation mechanism. Instead, it may stem from detrimental coupling effects during the battery's operational period [18]. Due to accuracy and computational complexity challenges, most existing remaining useful life (RUL) and health prediction models focus on singular degradation effect and usually ignore the integrated deterioration mechanisms that are normally involved in the capacity fade of batteries associated with the inadequacy of current health estimation tools.

The above challenges necessitate a robust and reliable predictive framework for prognostics and health monitoring (PHM) under a complexly hostile working environment. PHM is a multifaceted discipline for evaluating the extent of deviation or degradation of the system and is intended to detect incipient components or system faults [19]. Traditionally, PHM techniques use either a model-based (or physics-based) approach, data-driven approach or hybrid method [20-22]. In model-based approaches, mathematical equations, such as differential equations, are employed to represent the system. Also, the constitutive parameters or coefficients in the differential operators must be known *a priori*. Then, statistical estimation tools relying on parity relations are used to discover, isolate, and potentially predict degradation mechanisms [21, 23]. Numerical predictive models based on conservation principles are then used to make such prognostications. A variety of model-based methods have been proposed over recent years to predict the state of health (SOH), estimate the remaining useful life (RUL), and model the degradation mechanism of LIBs. For instance, Li et al. developed a universal capacity model based on the charging curve to estimate the SOH of LIBs, which inherits the advantages of the incremental capacity analysis method and avoids the data preprocessing procedure [24]. Safari et al. employed a multimodal physics-based aging model based on the continuous and small-scale growth of an SEI layer on the active anode particles' surface, highlighting the importance of accurate degradation modeling of cathodes [25]. Santhanagopalan and White proposed a model-based method based on an electrochemical model to estimate significant physical parameters that can ultimately measure the cell performance of LIBs with high accuracy [26]. Hu et al. proposed an integrated approach for the RUL prediction and capacity estimation of LIBs using the Gauss–Hermite particle filter technique to project the capacity fade and the end-of-service (EOS) [27].

On the contrary, data-driven approaches use statistical pattern recognition tools to detect anomalies in parameter data, isolate faults, and estimate a product's RUL [28, 29]. Generally, data-driven methods do not rely on the product-specific knowledge of the material parametersand failure mechanisms; in contrast, constitutive parameters of the product is evaluated by in situ monitoring and the anamoly in the system is detected without the need to know the failure modes [29]. Lately, numerous data-driven approaches have been proposed, such as Artificial Neural Networks (ANN) [30], Support Vector Machine (SVM) [31], Gaussian Process Regression (GPR) [32], Particle filter (PF), and fuzzy logic [28]. For instance, He et al. estimated the SOH, capacity degradation and RUL of LIBs using the Dempster–Shafer theory (DST) and the Bayesian Monte

Carlo (BMC) method [14]. Miao et al. employed the unscented particle filter (UPF) method to assess the batteries RUL and build the degradation model with small marginal error [33]. Wu et al. used an online method based on feed-forward neural network (FFNN) and Importance Sampling (IS) to estimate the LIBs remaining useful life [30]. Zhao et al. proposed a new approach in which they combined fault diagnosis results with statistical methods to construct a reliable battery model [34]. Despite the recent advances in developing data-driven methods, such techniques are still facing major challenges. For instance, LIB cells' degradation mechanism is extremely sensitive to the operating conditions. Hence, identifying the aging mechanism in a condition different from the training dataset becomes difficult. Also, different battery testing systems (BTS) have been utilized to monitor the state of voltage and current of batteries for health and RUL estimation. However, due to poor equipment accuracy in addition to the inability of the models to consider the noise influence, the performance of BTS could easily deviate and debilitate [35]. Moreover, current data-driven algorithms easily neglect the importance of proper tuning of hyperparameters and functions. When too many hyperparameters are adjusted simultaneously, rendering conclusive solutions becomes hard to achieve [36]. Additionally, developing robust and computationally efficient reduced-order models with high precision is always a demand that can monitor crucial safety-related side reactions such as lithium stripping or plating in real-time. Thus, developing multi-physics coupling models that have the ability to precisely describe the internal mechanism variations becomes a constant need [37]. These challenges ultimately complicate the battery packs health diagnosis and prognostics capabilities. Even though recent advances in artificial intelligence and deep learning algorithms have proposed practical solutions to these issues such as deep neural networks and generative adversarial networks that suggest more accurate non-linear fitting, to the best of authors knowledge, no research work has been conducted to utilize these types of methods for health diagnosis of battery pack. Therefore, it is crucial to introduce new methods that combine accurate aging prediction mechanisms with online health estimation that simultaneously consider the laboratory and modeling conditions [38].

In this context, this study aims at proposing and developing a novel data-driven approach called data-driven prognosis (DDP) that estimates the relevant constitutive parameters of the system *in situ* and detects deviations from the degradation dynamics of the system. The proposed DDP approach, circumvents the need for offline measurements or training and relies solely on in situ measurements. It considers the deviation parameters between the laboratory conditions used to develop the models and real operating conditions. The DDP method uses a deterministic framework, however, the solution's stochastic nature arises naturally from the underlying presumptions of the model regarding the order of the conservation principle and the number of dimensions involved. This approach can predict the onset of instabilities and identify the governing criteria that result in the battery pack's failure. It can be easily employed for any system in which such a priori testing is difficult or even impossible to conduct due to circumventing the need for offline testing or training. The instability forecasting capability of the DDP method has been verified to be accurate previously on a balloon burst experiment [39]. In this study, the DDP method was employed to analyze the health status of Li-ion batteries and predict when is the most probable time that the system reaches global instability and ultimately failure. The remainder of the paper is organized as follows: the DDP model development, including the algorithms introduced in section 2. Section 3 briefly details the experimental program, followed by results in section 4 and the conclusion in section 5.

## 2.0 Model Development

To understand the proposed data-driven prognostic scheme, it is attempted to characterize its concept by describing its essential components. For this purpose, first the idea behind the pairwise score is discussed, then the conservation principle is elaborated followed by the derivation of the length scale. In the field of geometric voting theories, numerous studies have been conducted to eliminate the conflict between the pairwise methods that present the societal outcome of elections. In particular, Saari [40, 41] has dedicated a substantial number of research studies to address this issue. The conflict originates from a special bloc of voters hidden in the given electorate and is described as a Condorcet profile [42]. A general solution to this conflict was proposed by Chandra [43] in which a reciprocal coordination technique was devised to remove the conflict in the pairwise electorates. The method assumes a governing conservation principle for the entire problem and then attempts to optimize the scores using the Reciprocal Coordination technique [44]. The proposed algorithm is described by the equation below:

$$\boldsymbol{a_{ij}^1 = a_{ij}^0 + CF \cdot \sum_{k \neq i,j}(a_{ik}^0 + a_{kj}^0), \forall i, j} \qquad (1)$$

Where $a_{ij}^0$ denotes the given initial pairwise score differences between i,j, $a_{ij}^1$ represents the revised pairwise score differences after one iteration defined late on and CF is a positive constant described as the confidence factor.

In the proposed approach, it is assumed that a conservation principle is applied to the observation points, which may be the conservation of energy, conservation of linear and angular momentums, or all of the above. For a voting problem, this may be the simple fact that a total number of votes be constant. In the proposed DDP method, it is assumed that the total momentum in the system is conserved. Following such a logic, a dimensionless form of length scale value containing the correct combination of geometric and material parameters, is derived from the conservation functional. The length scale values are then used to capture the magnitude of curvature in the system as a form of absorption or release. To do so, suppose we have an observable body or phenomena with finite number of observation points. At each of these observation points, information is collected at discrete instants of time in multiple dimensions. These points are collectively considered, and a normalized relationship is constructed for each pair of points for the entire set of observation points. Calling the two observation points A and B, at a specific dimension d, the normalized relationship, $a_d^{AB}$, is defined as equation (1):

$$a_d^{AB} = \frac{u_d^A - u_d^B}{u_d^A + u_d^B + 2\bar{m}_d} \tag{2}$$

$u_d^A$ and $u_d^B$ are the measured values at dimension d at points A and B. The parameter $\bar{m}_d$ is a small constant, to be determined later. The purpose of such a normalized relationship is to have an understanding of the behavior of points in a pairwise manner. Next, a governing conservation principle is applied at each point that must be satisfied at all times, considering that the system remains in a conservative state. The principle of conservation of linear momentum is chosen for this purpose. Furthermore, it will be attempted to satisfy the three canonical requirements: equilibrium, compatibility, and constitutive laws. To develop such a model, as the first attempt, a piecewise second-order utility function is assumed to sufficiently describe the system's interactions. However, the nature of the piecewise quadratic potential function can differ from one point to the other. Additionally, it is assumed that the objectivity in the system is also ensured at all times implying that the state of observed system remains unchanged due to rigid body transformations of the reference frame. Enforcing these requirements, the system is described by considering the conservation of linear momentum in the neighborhood of point A as [39, 45]:

$$R_i^A - \beta_{ik}^A \times \Delta H_k^A = 0 \qquad (3)$$

$R_i$ represents the rank (long-term rank) in dimension i at point A. $H_k$ represents the Borda count (short term rank) at dimension k at point A. $\Delta H_k$ represents the change in the Borda count in dimension k during a time step. Borda count in this context, is calculated by constructing a global skew-symmetric matrix that represents the normalized value for each pair of points and summing them at each row or column. Long-term rank is similarly computed by performing one round of iteration using equation (1). For more details about these parameters, readers are encouraged to refer to [40, 41, 43, 44]. The parameter $\beta_{ik}$ is a non-dimensional quantity that represents a second order norm of the length scale around a point, which is described as:

$$\beta_{ik} = \left(\frac{1}{\Delta t^2}\right) \times \left(\frac{\rho}{E_{ijkl}}\right) \times \|L_l L_j\| \qquad (4)$$

Here, $\Delta t$ is the change in time step, $\rho$ is the density and $E_{ijkl}$ is the tangent modulus. Substituting the above expressions into equation (2), we obtain the general format for the length scale that is expressed as:

$$E_{ijkl} \left[\frac{R_i}{L_l L_j}\right] = \frac{\rho}{(\Delta t)^2} \times \Delta H_k \qquad (5)$$

Length scale is defined as the region of validity along a particular dimension in which the linearization of the governing equation is valid. Furthermore, the conservation of angular momentum and the symmetry of the potential function about an interchange of the state variables are enforced. This is administered by requiring interchangeability of (i and j), (k and l) as well as (i,j and k,l) as pairs. Existence of symmetrical potential function, is also mandated by the work conjugacy requirement of stress and strain [46]. After enforcing all the symmetry requirements, a general expression of the potential function at each observation point is expressed as:

$$\frac{E_{ijkl}}{2} \times \left[\frac{R_i}{L_l L_j} + \frac{R_j}{L_k L_i} + \frac{R_k}{L_j L_l} + \frac{R_l}{L_i L_k}\right] = \frac{\rho}{(\Delta t)^2} \times \left[\Delta H_i + \Delta H_j + \Delta H_k + \Delta H_l\right] \tag{6}$$

The number of solutions for the potential function is dependent upon the number of dimensions involved in the system. In general, the number of solution, also called the number of roots, is $2^{\text{dimsize}}$ (dimension size). At each point there are number of dimensions; the root at each point in each dimension corresponds the roots at every other dimension. Thus at each point at each dimension, the number of equations is equal to the number of dimensions. For instance, if there are 4 dimensions at a specific point, the number of equations are 4. The roots then can be obtained by solving the set of (d x d) simultaneous equations at each observation point. Thus, the proposed strategy requires solving N matrices (when N is number of observation points) each with dimension d, resulting in a very computationally efficient and massively parallelizable scheme.

The component $\bar{m}_d$ essentially sets the datum, and the coordinate of the origin is set at $-\bar{m}_d$ in the corresponding dimension. Owing to the fact that only a least square approximation is used for calcualting $\bar{m}_d$, it is applied universally to all points (with the purpose of setting a same datum for all observation points), however, the values of $\bar{m}_d$ in different dimensions are different and are set independently.

Next, the obtained values of volumetric and shear length scales are used to compute the shear and volumetric curvatures. This is achieved by defining a composite curvature as (Equation 7):

$$\kappa = \alpha\kappa_v + (1-\alpha)\kappa_s \tag{7}$$

Where, $\kappa_v$ and $\kappa_s$ are volumetric and shear curvatures, respectively, that are calculated as:

$$\kappa_v = \frac{(R \times \bar{L}_v - H \times \tilde{\bar{L}}_v)}{(((\bar{L}_v^2) \times \tilde{\bar{L}}_v) \times (1 + (H/\bar{L}_v)^2)^{1.5})} \tag{8}$$

$$\kappa_s = \frac{(R \times \bar{L}_s - H \times \tilde{\bar{L}}_s)}{(((\bar{L}_s^2) \times \tilde{\bar{L}}_s) \times (1 + (H/\bar{L}_s)^2)^{1.5})} \tag{9}$$

$\tilde{\bar{L}}$ represents the homogenized version of $\bar{L}$, $\alpha$ is the proportion by which volumetric curvature takes place and R and H are rank and Borda count that were described previously. Substituting the volumetric and shear curvature values in equation (7), it is attempted to minimize the composite curvature by setting the LHS of the equation to zero and obtain the $\alpha$ values at each point across each dimension. Then, the mean of all squared $\alpha$ values are taken and used in the equation (6) again using that unique value to obtain the composite curvature.

To identify a specific range for the alpha values, we employ the composite stress equation that is described by [46]:

$$\sigma_{ij} = \lambda\varepsilon'\delta_{ij} + 2G\left(\varepsilon_{ij} - \frac{1}{3}\varepsilon'\sigma_{ij}\right) \tag{10}$$

Where $\lambda$ and G are bulk and shear modulus, respectively and are defined as:

$$\lambda = \frac{vE}{(1+v)(1-2v)} \tag{11}$$

$$G = \frac{E}{2(1+v)} \tag{12}$$

By comparing equation (10) with equation (7), a similarity is noticeable between these two equations and their volumetric and shear components. The range of proportionality is obtained from this resemblance and considering the proportions of:

$$\frac{\lambda}{2G} = \frac{\alpha}{1-\alpha} = \frac{\nu}{(1-2\nu)} \qquad (13)$$

We already are aware of the range of Poisson's ratio that is in the range of $-1 < \nu < 0.5$. hence, the range of α should be between $-0.5 < \alpha < 1$. By computing alpha, the mode mixity between the dilatational and shear modes in the problem is assigned.

## 2.0 Post processing [39, 45]

The proposed method relies on a set of criteria that evaluates the magnitude of instability in the system. These criteria uses the information that were obtained from length scale and curvature calculations. They are discussed in the following sections.

### 2. 1 Categorizing the instability of observation points

The next step in model development is the assignment of the Path Dependency Index (PDI) to the observation points following the determination of curvature and associated length scales. The PDIs are divided into nine possible categories each representing a magnitude of instability, in which category 1 represents no instability and category 9 denotes complete instability (Table 1). All 16 number of roots (in 4D) for the length scale obtained from the previous step have to be examined individually for this stage. The severity of instability will be measured based on the ratio of absolute value of the obtained composite curvature equation divided by the absolute value of curvature that is determined from the reciprocal of length scales, i.e. $\bar{L}$ and $\tilde{\bar{L}}$; we call them kappa short and kappa long, respectively. If the obtained value is less than one in both cases, we conjecture that the observation point does not exhibit short or long-term instability. Depending on whether the point reaches short-term or long-term instability, it will be categorized based on Table 1. The initial 7 categories are related to PDI, and the remaining 2 categories, i.e. categories 8 and 9, are designed for further post-processing, which depicts the global transcendation index or GTI. Suppose the observation point falls in the category 5 or more. In that case, we conclude that the point enters the path dependency stage and might illustrate local instability with the potential to enter the global instability phase.

**Table 1. Path dependency index (PDI) categories**

| Category | Definition of the Category |
|---|---|
| 1 | Complete stability: (abs(*k)/abs(k-long)) <1 and (abs(k)/abs(k-short)) <1 |
| 2 | Long-term stability & short term instability: (abs( k)/abs( k-short))>1 |
| 3 | Long-term instability & short term stability: (abs( k )/abs( k-long))>1 |
| 4 | Provisional stability: Changing mode mixity might cause local instability |
| 5 | Short-term and long-term instability: local instability cannot be be controlled by altering mode mixity |
| 6 | Short-term and long-term instability in dimension > 1 |
| 7 | Short-term and long-term instability in dimensions > 2 |
| 8 | Chain length > short-term critical chain length, PDI > 5 |
| 9 | Chain length > long-term critical chain length, PDI > 5 |

* k = kappa (curvature)

## 2. 2 Determination of chain length

Following the assignment of PDIs, it is necessary to check how long a defect or a chain of possible unstable points might continue in either direction in the order of their ranks, called chain length. This calculation is done to facilitate the determination of chain length in abstract systems such as genetics or economic systems that is associated with the constitution of energy exchange pathways. For a point to enter an energy exchange pathway, it must exhibit a $PDI \geq 5$. This threshold is identified as the critical gateway to transcendation to the next aggregated scale in the hierarchy and might enter to global transcendation phase.

## 2. 3 Global transcendation index

After the system enters the path dependency phase, if it meets two additional measurements, it is conjectured that it is progressing toward failure. These measurements are: (1) the locally path-dependent observation points form a chain, whose length exceeds a critical threshold value, (2) the aggregated level of the system exceeds a critical threshold called residual curvature. Upon meeting these measurements, the system enters the global transcendation stage and the category 8 is assigned to the point if it reaches $PDI > 5$, and $GTI > 0$ in one dimension, and category 9 is assigned if both of these conditions are satisfied in multiple dimensions.

## 2. 4 Aggregation or zoom-out procedure

The aggregation procedure (or zoom-out plot) is followed and calculated to determine the critical chain length and residual curvature (also called aggregated curvature) of the system for each instant of time. The residual curvature provides a measure of the potential energy exchange rate of the system with its surrounding environment. According to the definition of zoom out procedure, the observed curvature for a conservative system should be approximately zero in instances when the stand-off distance of the observer is both zero and infinity. Such a procedure serves as a mean to identify the critical length of the energy exchange pathway that is required for the local instabilities to transcend to global scales. The energy exchange rate also needs to exceed a critical threshold value to meet the sufficiency conditions for transcendence of local instabilities to global scale. Because the critical energy release rate is related to the system's constitutive property, which is determined normally by conducting offline testing, the proposed method assumes that almost all of the energy stored along the exchange pathways gets released when the local instability transition to a global scale. Hence, the energy rate fluctuates rapidly and falls nearly to zero during such a transcendent procedure. This rapid fluctuation in the energy exchange rate is used in this method to account for a trigger initiated every time the energy rate drops more than 80% during a step (which is arbitrarily chosen and associated with the inherent noise and computational burdens). Therefore, a point reaches the global transcendent stage following exceeding a critical chain length and 80% or more drop in the energy release rate during a single time step, constituting a GTI greater than 0.

The failure prediction relies on three primary measurements, and for a system to initiate instability and it needs to meet the following metrics: (1) the system enters the path dependency phase, (2) long-term chain formation is triggered, and (3) residual curvature (or energy release rate) is dropped by 80% or more. A system is said to approach to failure if it meets these conditions simultaneously and the prognosis scheme makes a prediction based on the instant of time that these measurements were met. It should be noted that the method was previously validated based on the studies conducted in [39, 45].

## 3.0 Methods

The experiment done were mainly aged battery cycling tests. Since the proposed DDP requires large amount of data (i.e., charge capacity, discharge capacity, current, and voltage) at each data collection and sampling to analyze, we selected commercially available rechargeable lithium batteries that are naturally aged in our lab and quickly failed during testing. Cost and equipment limiting factors led to the decision to use Kokam superior lithium polymer pouch-sized batteries that are 95 mm in height and 3.5 mm in depth and 64 mm wide. As suggested by the manufacturer, the maximum charge current and voltage were set to below 2020mA and 4.2V; the maximum discharge current and voltage were 3030 mA with a cut-off voltage of 3.0V. The battery-rated capacity, nominal voltage, and cycle life are 2100 mAh, 3.7V, and 500 cycles (@ 0.5C charge and discharge). Data collection used a Neware BTS5V6A8CH battery testing system. Constant Current Constant Voltage (CCCV) charging and discharging protocol was adopted for all tests. The test included loading and unloading the cells to their cut-off voltage and subsequently charging them to their maximum safe voltage (100% SOC). Then, the Vencon recharges the cells to 30% SOC in order for safe storage. Furthermore, each step is performed based on the battery specifications and desired test (for instance, maximum C-rating).

## 4.0 Results and Discussion

After the batteries were analyzed, the results of the experimental tests were extracted. The battery data that were extracted are voltage, current, charge, and discharge capacity. Figures 1 & 2 show the variation of peak charge and discharge capacity for the battery cycling in 48 days (i.e., 48D) @1C and 54 days (i.e., 54D) @2C.

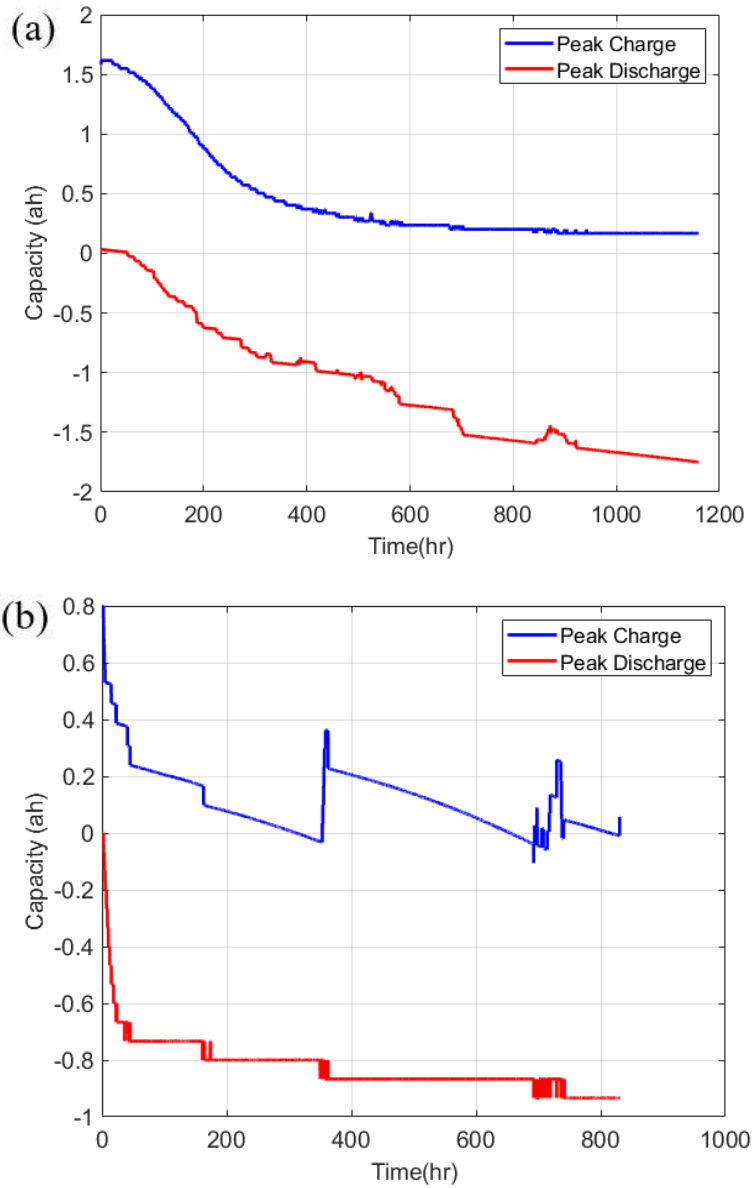

**Figure 1. Experimental result of the battery charge discharge capacity: a) 48D battery cycling @1C b) 54D battery cycling at @2C**

## 4.1 Path dependency index (PDI)

As described in the model development section, first the long-term and short-term ranks are computed using the constructed global matrix and equations (1) & (2) at each instant of time. Using the two obtained values, the conservation principle is enforced at each point and the dimensionless length scales (dilatational and distortional) using the given Poisson's ratios are obtained, allowing for the calculation of local curvature. The obtained local curvature information is then used to compare the curvature values with the critical threshold (denoted by the reciprocal of the length scale). Based on the comparison results, each point at a specific root is assigned a PDI category based on Table 1 to evaluate the amount of instability that occurs at each point. A PDI greater than 5 represents local instability in the system with the potential to transcend to global instability. Next, consistent with the assumption of piecewise second-order utility function around each observation, the zoom-out aggregation procedure is followed. This procedure is done to identify the critical chain length in the system defined as the magnitude of energy exchange pathway required for a local instability to transcend to global scales. Coupling occurrence of chain length greater than the critical value in addition to the exceedance of energy exchange rate through such a pathway, constitutes the global instability in the system. Such critical energy release rate is an inherent property that is normally calculated through offline testing. In this work, instead of such offline measurement, it is conjectured that almost all of the energy stored along such a pathway gets released when the local instabilities transition to global scales. Hence, during such transcendent, the energy exchange rate fluctuates rapidly and reaches zero. This rapid energy exchange rate is used in the scheme as the instant that energy drops greater than 80% at a time instant. Selection of such a rate is arbitrary in this method alluding to the inherent noise floor in the procedures. The results of the PDI categorization are shown in Figure 2. The figure demonstrates the percentage of points reaching a specific PDI for different time instants. Based on the results of PDI calculation, the system enters path dependency phase in its initial stages.

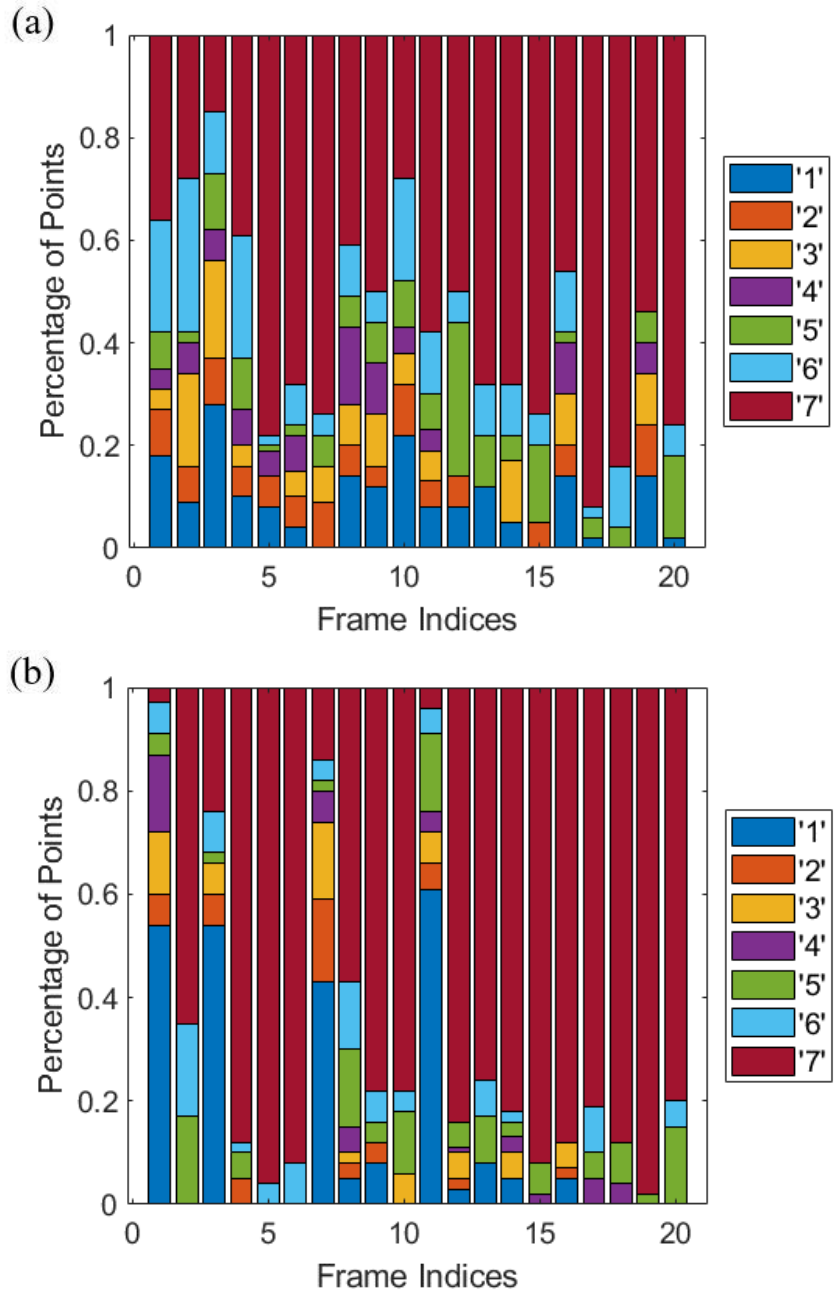

**Figure 2. Path dependency index for peak charge capacity: a) 48D battery, b) 54D battery**

The figure illustrates the number of roots that have a PDI category larger than 5 only without the chain length information considered. The times when the roots are equal to and more than 50% in number of points, is said to path dependent. The system is said to be path-dependent if at least one point in the observation body is path-dependent. Figure 3 shows the plots that highlight the moments when the system might have become path-dependent only. In other words, they show the number of roots that have a PDI category greater than 5 only without the chain length information. Based on the results, the systems in both cases enter the path dependency at their early stage.

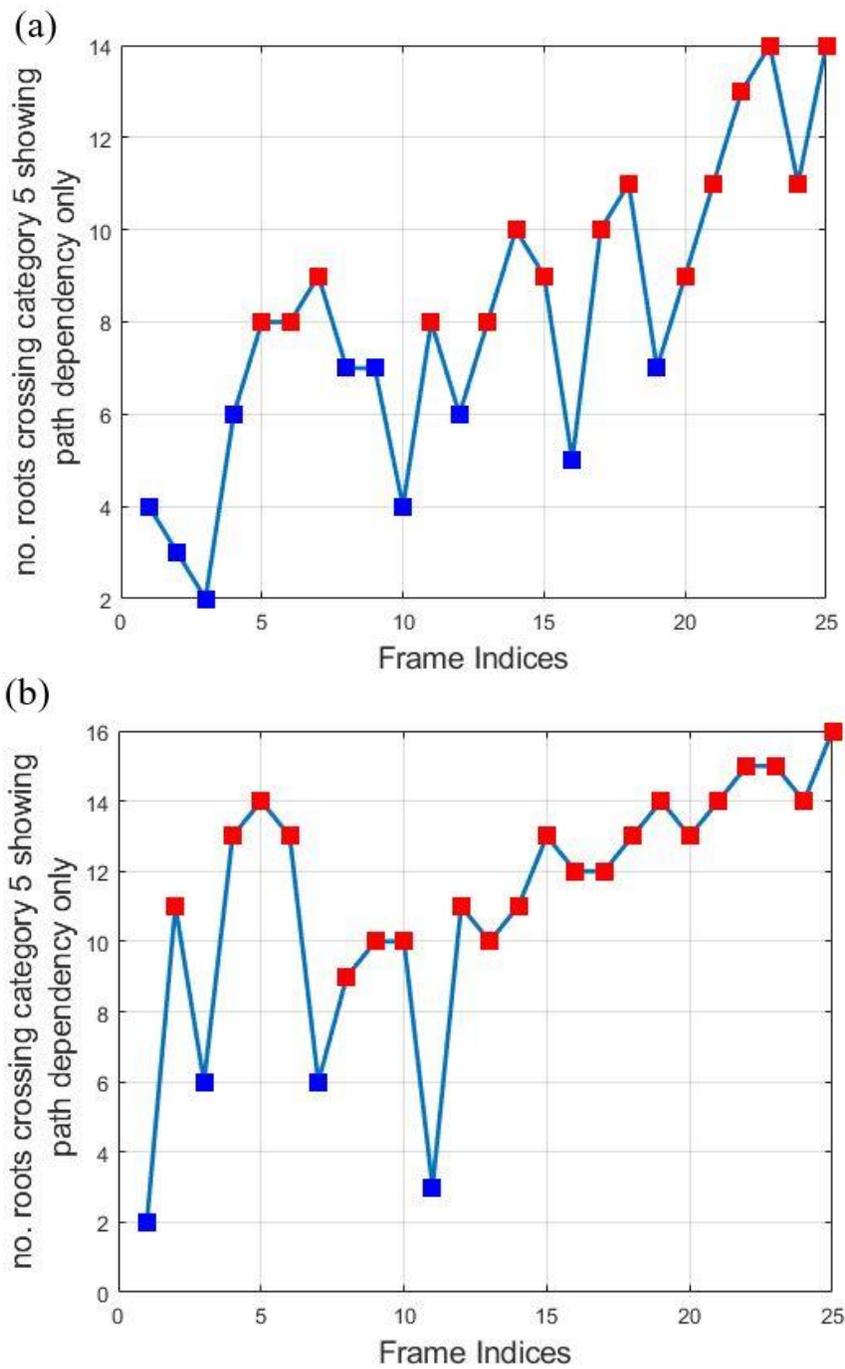

**Figure 3. Number of roots having category greater than 5 showing path dependencies only in peak charge: a) 48D battery, b) 54D battery**

## 4.2 Chain length formation

Figure 4 illustrates the number of roots that had PDI ≥ 5 and also crossed the critical chain length threshold (22 in the battery system). The value of critical chain length is calculated from the zoom-out plot which is shown in Figure 6, as a sample plot for an arbitrary dimension. After calculating the curvature values (κ or kappa) as well as short-term and long-term critical values which are denoted by ($\frac{1}{LBar}$, $\frac{1}{LBarTilda}$ respectively) at every aggregation level, the kappa graph is extrapolated in its opposite region using the symmetry condition. Then, $\frac{1}{LBarTilda}$ line is extended backwards until it intersects with the mirror line (around y-axis) of kappa graph. Then, the logarithmic value of this intersection is identified and considered as the number of points related to the aggregated frame and critical chain length value. Such a value provides an estimate for long-term critical chain length. Critical chain length is calculated to determine how long a defect continues in each direction in order of their ranks. The critical chain length depends on the number of chosen observation points. The instants of time when the roots are equal and greater than 50% (8 out of 16) in number, the system is said to be near failure in all those times. Based on the results, for both 48D and 54D batteries, the systems cross the long-term localization index at an early stage and this phenomenon is more severe for 54D batteries.

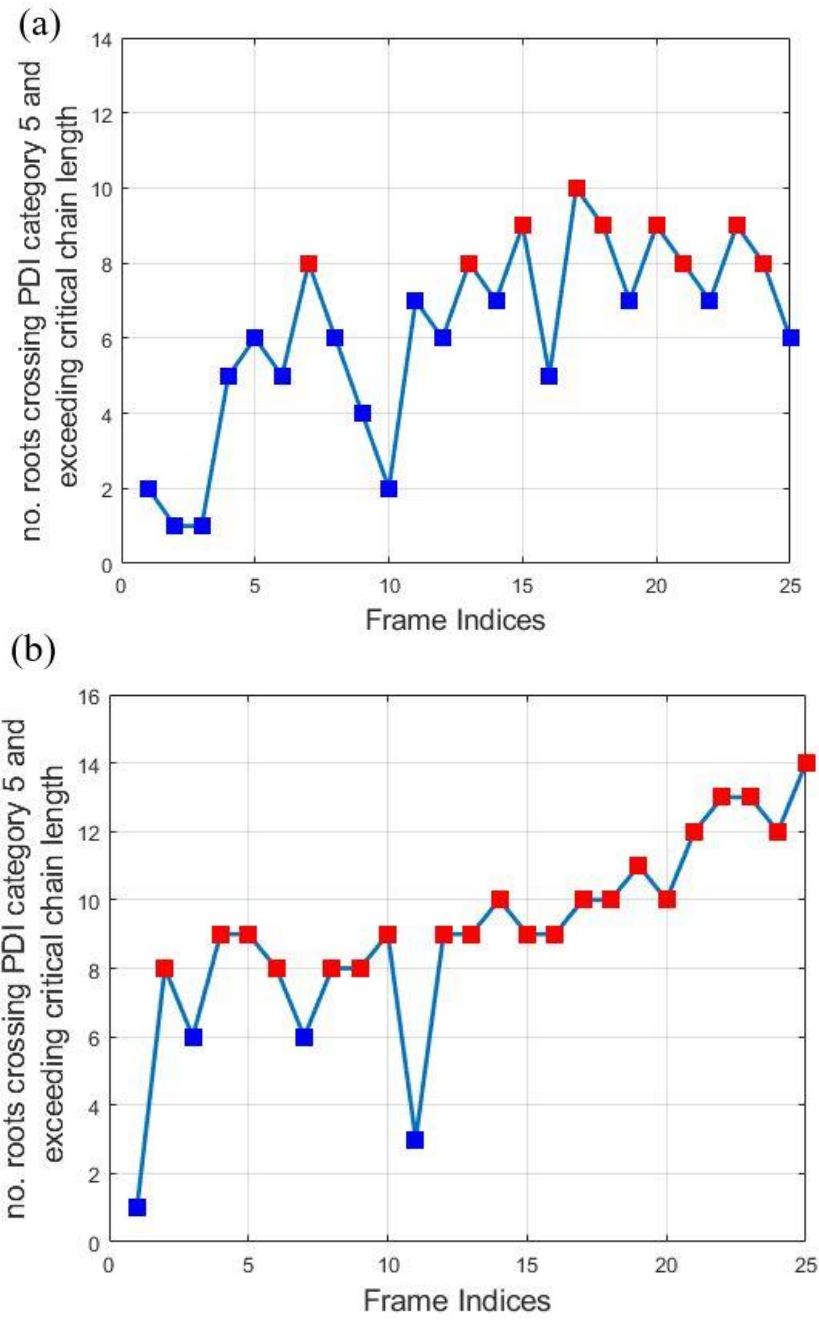

**Figure 4. Number of roots having category greater than 5 and having chain length greater than number of points in peak charge capacity: a) 48D battery, b) 54D battery**

**4.3 Residual curvature**

As discussed in previous sections, Figure 3 illustrates when the system is only path-dependent, while Figure 4 illustrates when local instability might transcend to global instability by forming an energy exchange pathway longer than the critical threshold value. The residual curvature results for the systems at different time indices are shown in Figure 5. The residual curvature concept is used herein to provide a dimensionless aggregated measure of the magnitude of the energy exchange rate for the system. According to this concept, if the residual curvature surpasses a critical threshold value, the local instability can transcend to global instability in the system.

Here, another fact was used is that the stored energy in the system is depleted rapidly during such global transcendation. Thus, the rate of change, such rapid drop, is constantly monitored to detect any imminent failure of the system. At time instances in which this rapid drop exceeds more than 80%, it is said that the system is progressing toward failure. After each time instant was analyzed, the results were obtained and shown in Figure 5, which shows the percentage of a total number of roots (out of 64) in all four dimensions of the battery that had a drop of residual curvature larger than 80%. The battery will dissipate a large amount of energy if the number of roots reaches a value larger than 20% (12 out of 64 roots)

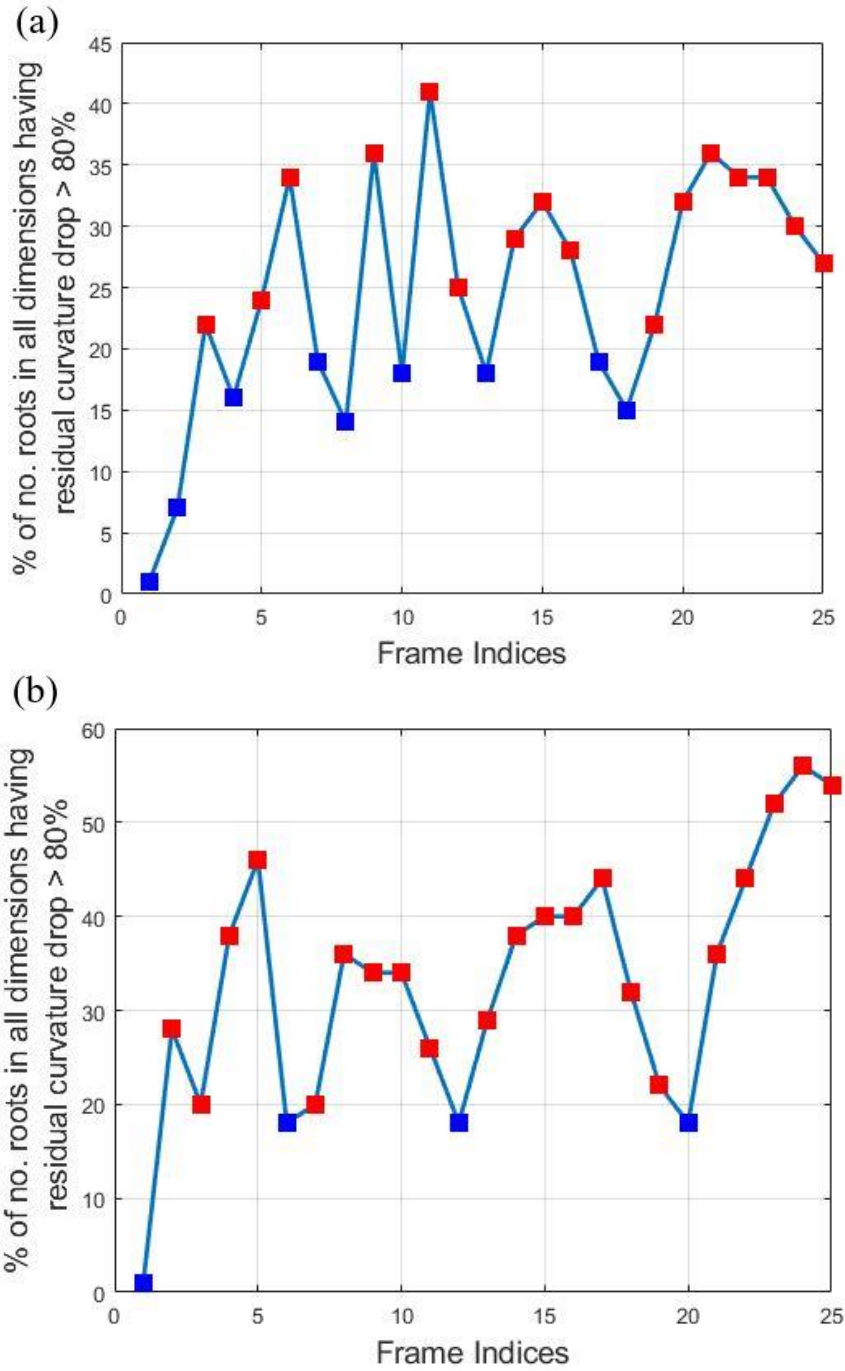

**Figure 5. Number of roots having residual curvature drops more than 80% in peak charge capacity: a) 48D battery, b) 54D battery**

## 4.4 Critical localization index

The Zoom out plots, as stated previously, are used to calculate the critical chain length in the system. Based on the results of the critical chain length calculation shown in Figure 6 for 48D battery in an arbitrary dimension, the number for the critical chain length can be obtained and converted to a number corresponding to the number of observation points were considered for the lithium-ion batteries. The number derived from Figure 6 represents the critical chain length for 9 points and later converted to match the number of points considered for the system. This number is considered for the chain length algorithm to determine how many points form a chain that is larger than the calculated value.

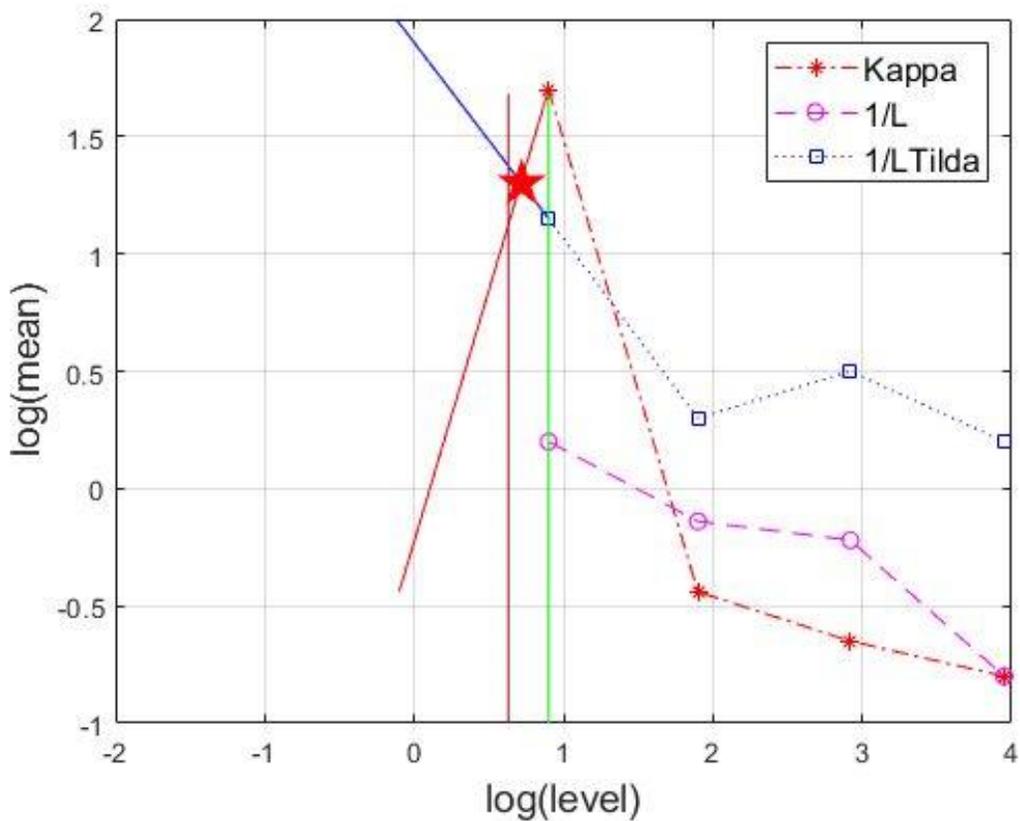

**Figure 6. Zoom out plot – Chain length calculation in peak charge capacity (48D battery)**

## 4.5 Composite failure prediction

The final failure prediction takes all the above analysis into account in order to make a prognostication about the system as to when is the most probable time that it fails. The results are shown for 48D and 54D batteries in Figures 7 and 8. To prognosticate the failure, the system first needs to enter the path dependency phase. Then, if the chain formed at each time instant exceeds the critical chain length in both short-term and long-term phases, the dislocations have the potential to form a line defect which can cause failure in the system.

Furthermore, the system needs to have greater energy absorption or release than the critical energy exchange rate. Such activity is seen through large and rapid drops in residual curvature. A drop of larger than 80% is used here as a trigger in the system. Once both of these criteria are met (GTI>0) after the system entered the path dependency mode (PDI>5), failure can be predicted for the system. In the battery prognostics, the systems were under constant monitoring when it entered the path dependency stage. Based on the results, for the 48D battery, the system is supposed to fail at time instant 15, while for the 54D battery, the system is expected to fail at time instant 5.

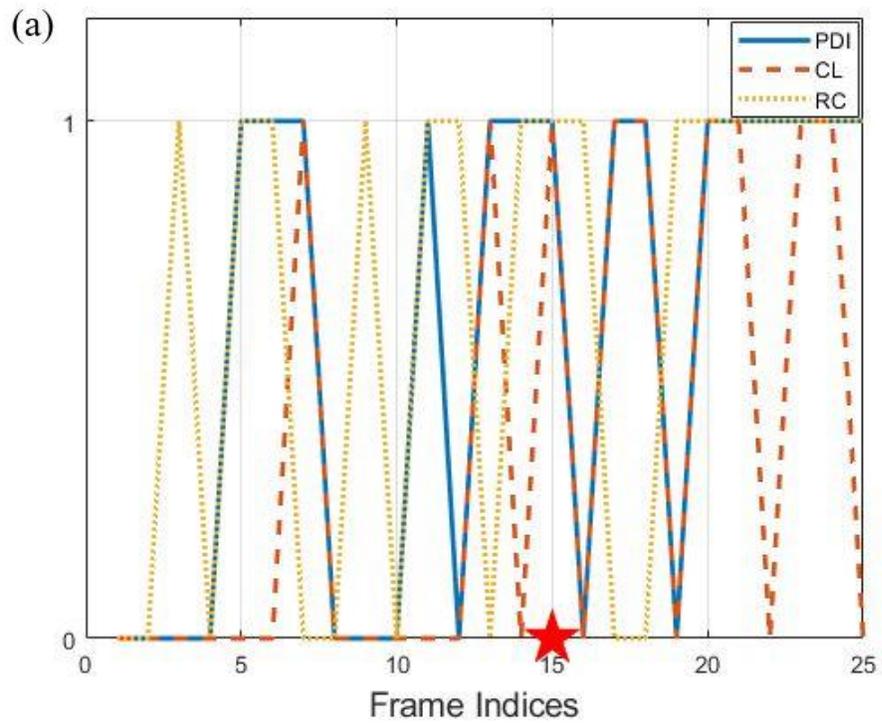

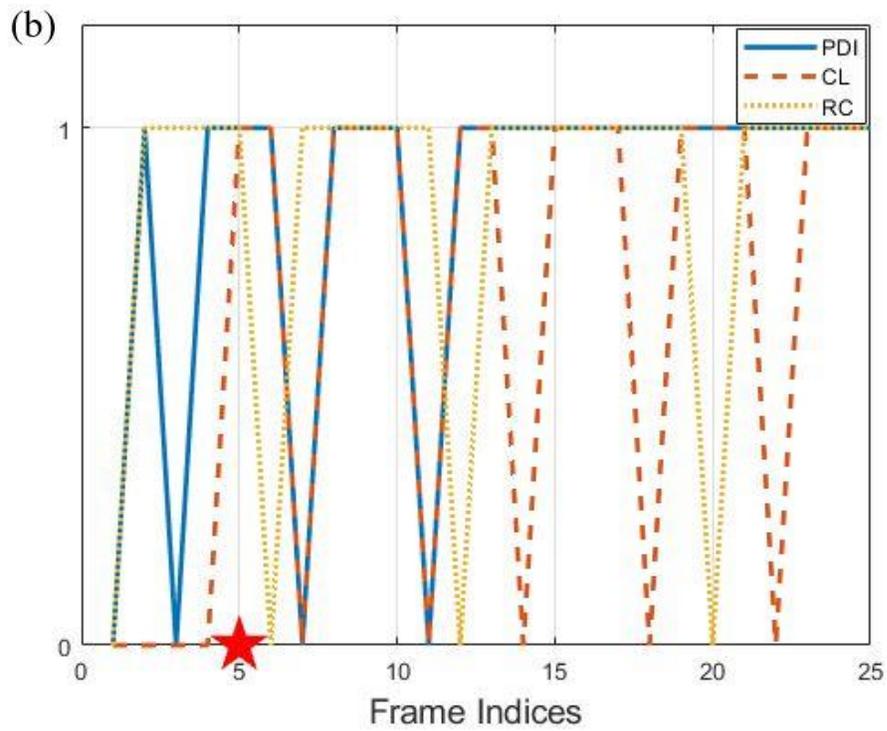

**Figure 7. Composite failure prediction for peak charge capacity: a) 48D battery, b) 54D battery**

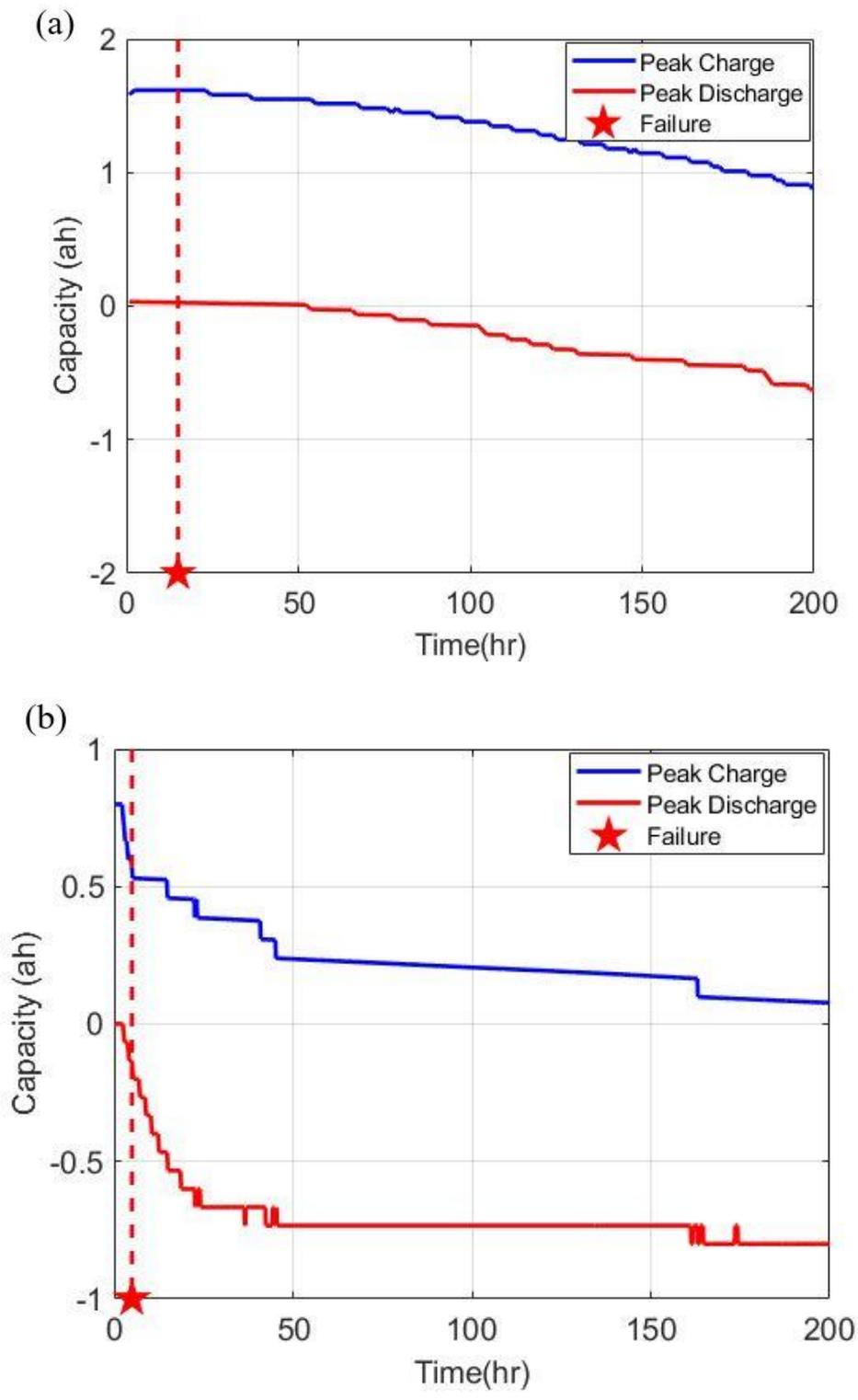

**Figure 8. Composite failure prediction on experimental graph: a) 48D battery, b) 54D battery**

## 5.0 Conclusion

One of the main obstacles in the further adaptation of the current electric vehicle market is the reliability and safety of battery packs. A universal solution for safe operation and accurate monitoring of batteries is the implementation of PHM frameworks. This study proposed and employed a novel data-driven method to analyze the health status of Li-ion batteries. The method is suitable for handling multi-scale and multi-physics problems that is based on data gathered in a continuous process. Unlike many PHM methodologies that rely on the system's available information or constitutive parameters obtained by offline testing, the proposed scheme obviates the need for a priori offline testing.

Furthermore, most models neglect important governing state parameters of the systems in order to have a simplified and reduced-order model, which can lead to inaccurate estimation of the health and status of battery packs. Based on the results of this study, it has been shown that the proposed method was able to detect fault and anomalies in the system in the form of excessive curvature. The excessive curvature was extracted by enforcing a local conservation principle around each observation point. Using such an approach, allowed for a systematic assessment of the system's health and status. Due to the broad applicability of the proposed method, it can be used for a variety of systems without the need to include any pre-existing information about that system.

## 6.0 Acknowledgements

L.L. appreciates the support from the National Science Foundation under Award #1840732, KS NASA EPSCoR Program, Kansas Soybean Commission, KU RISe Award, KU Research GO awards, and KU General Research Funds. Any opinions, findings, and conclusions orPage 31 of 37 recommendations expressed in this material are those of the author(s) and do not necessarily reflect the views of the funding agencies.

## 7.0 References

1. EDTA, *Electric Drive Transportation Association (EDTA), Washington Dc.* 2021.
2. BloombergNEF, *Electric Vehicle Outlook.* BloombergNEF (BNEF), 2021.
3. IEA, *Global EV Outlook 2020: Entering the decade of electric drive?* . International Energy Agency (IEA), Technology report, 2020.
4. Lu LG, e.a., *A review on the key issues for lithium-ion battery management in electric vehicles.* J Power Sources 2013. **226**: p. 272-288.
5. K. Forrest, e.a., *Estimating the technical feasibility of fuel cell and battery electric vehicles for the medium and heavy duty sectors in California.* J Applied Energy, 2020. **276**.
6. Pinson MB, B.M., *Theory of SEI formation in rechargeable batteries: capacity fade, accelerated aging and lifetime prediction.* J Electrochem Soc 2012. **160**: p. A243-50.
7. L. Liu, J.P., X. Lin, A. M. Sastry, W. Lu, *A thermal-electrochemical model that gives spatial-dependent growth of solid electrolyte interphase in a Li-ion battery.* J Power Sources 2014. **268**: p. 482-490.
8. P. Guan, L.L., X. Lin, *Simulation and Experiment on Solid Electrolyte Interphase (SEI) Morphology Evolution and Lithium-Ion Diffusion.* J. Electrochem. Soc, 2015. **162**: p. A1798-A1808.

9. U.R. Koleti, T.Q.D., J. Marco, *A new on-line method for lithium plating detection in lithium-ion batteries.* J Power Sources, 2020. **451**: p. 227798.

10. B. L. D. Rinkel, D.S.H., I. Temprano, C. P. Grey, *Electrolyte Oxidation Pathways in Lithium-Ion Batteries.* J American Chemical Society 2020. **142**: p. 15058-15074.

11. Birkl CR, R.M., McTurk E, Bruce PG, Howey DA, *Degradation diagnostics for lithium ion cells.* J Power Sources, 2017. **341**: p. 373-86.

12. Celina, M., *Lithium-ion batteries hazard and use assessment.* US: Springer, 2011.

13. L. Liu, M.Z., *Modeling of SEI layer growth and electrochemical impedance spectroscopy response using a thermal electrochemical model of Li-ion batteries* ECS Trans, 2014. **61**: p. 43-61.

14. He W, W.N., Osterman M, Pecht M, *Prognostics of lithium-ion batteriesbased on Dempster–Shafer theory and the Bayesian Monte Carlo method.* J Power Sources 2011. **196**: p. 10314-21.

15. Barré A, D.B., Grolleau S, Gérard M, Suard F, Riu D, *A review on lithiumion battery ageing mechanisms and estimations for automotive applications.* J Power Sources 2013. **241**: p. 680-9.

16. Li Y, V.M., Choi SS, Farrell TW, Tran NT, Teague J, *Development of a degradation-conscious physics-based lithium-ion battery model for use in power system planning studies.* Appl Energy 2019. **248**: p. 512-25.

17. A. Maheshwari, e.a., *Optimizing the operation of energy storage using a non-linear lithium-ion battery degradation model.* Appl Energy 2020. **261**: p. 114360.

18. X. Lin, J.P., L. Liu, Y. Lee, A.M. Sastry, W. Lu, *A comprehensive capacity fade model and analysis for li-ion batteries.* J Electrochem Soc, 2013. **160**: p. A1701-A1710.

19. S. Saxena, Y.X., M.G Pecht *PHM of Li-ion batteries Prognostics and health management of electronics* Wiley, 2018: p. 349-375.

20. S. Cheng, M.H.A., M.G. Pecht, *Sensor Systems for Prognostics and Health Management.* Sensors, 2010. **10**: p. 5774-5797.

21. J. Liu, D.D., K.A. Marko, J. Ni, *Mechanical Systems and Signal Processing.* 2009. **23**: p. 2488-2499.

22. Guo J, L.Z., Pecht M, *A Bayesian approach for Li-Ion battery capacity fade modeling and cycles to failure prognostics.* J Power Sources 2015. **281**: p. 173-84.

23. G.J. Vachtsevanos, F.L., A. Hess, B. Wu, *Intelligent fault diagnosis and prognosis for engineering systems.* Wiley Online Library, 2006.

24. Li X, J.J., Yi L, Chen D, Zhang Y, Zhang C, *A capacity model based on charging process for state of health estimation of lithium ion batteries.* Appl Energy, 2016. **177**: p. 537-43.

25. M. Safari, M.M., A. Teyssot, C. Delacourt, *Multimodal physics-based aging model for life prediction of Li-Ion batteries.* J. Electrochem. Soc, 2009. **156**: p. A145-A153.

26. S. Santhanagophlan, R.W., *State of charge estimation using an unscented filter for high power lithium ion cells.* Int. J. Energy Res, 2010. **34**: p. 152-163.

27. Hu C, J.G., Tamirisa P, Gorka T, *Method for estimating capacity and predicting remaining useful life of lithium-ion battery.* Appl Energy 2014. **126**: p. 182-9.

28. B. Saha, K.G., S. Poll, J. Christophersen, *Prognostics methods for battery health monitoring using a Bayesian framework.* IEEE Transactions on instrumentation and measurement, 2008. **58**: p. 291-296.

29. M, P., *Prognostics and health management of electronics.* New York, NY: Wiley Interscience, 2008.

30. Wu J, Z.C., Chen Z, *An online method for lithium-ion battery remaining useful life estimation using importance sampling and neural networks.* Appl Energy 2016. **173**: p. 134-40.

31. ME, T., *The Relevance Vector Machine: advances in neural information processing systems.* 2000: p. 652-8.

32. Rasmussen Carl Edward, W.C.K., *Gaussian processes for machine learning.* MIT Press, 2006.

33. Q. Miao, L.X., H. Cui, W. Liang, M. Pecht, *Remaining useful life prediction of lithium-ion battery with unscented particle filter technique.* Microelectron. Reliab 2013. **53**: p. 805-810.

34. Zhao Y, L.P., Wang Z, Zhang L, Hong J, *Fault and defect diagnosis of battery for electric vehicles based on big data analysis methods.* Appl Energy 2017. **207**: p. 351-62.

35. Xia, B., Cui, D., Sun, Z., Lao, Z., Zhang, R., Wang, W., Sun, W., Lai, Y., Wang, M, *State of charge estimation of lithium-ion batteries using optimized Levenberg- Marquardt wavelet neural network.* Energy 2018. **153**: p. 694-705.

36. M.S. Hossain Lipu, M.A.H., A. Hussain, A. Ayob, M. H.M. Saad, T.F. Karim, Dickson N.T, *How Data-driven state of charge estimation of lithium-ion batteries: Algorithms, implementation factors, limitations and future trends.* J Cleaner Production, 2020. **277**: p. 124110.

37. W. Xie, X.L., R. He, Y. Li, X. Gao, X. Li, Z. Peng, S. Feng, X. Feng, S. Yang, *Challenges and opportunities toward fast-charging of lithium-ion batteries.* J Energy Storage, 2020. **32**: p. 101837.

38. Y. Li, K.L., A.M. Foley, A. Zülke, M. Berecibar, E. Nanini-Maury, et al, *Data-driven health estimation and lifetime prediction of lithium-ion batteries: a review.* Renew Sustain Energy Rev, 2019. **113**: p. 109254.

39. Chandra A, K.O., *Data driven prognosis: a multi-physics approach verified via balloon burst experiment.* Proceedinga of the royal society A, 2015a. **471**(2176).

40. DG, S., *Explaining all three alternative voting outcomes.* J. Econ. Theory, 1999. **87**: p. 313-355.

41. DG, S., *Chaotic elections! A mathematician looks at voting.* Providence, RI: American Mathematical Society., 2001.

42. DG, S., *Mathematical structure of voting paradoxes I. Pairwise votes.* Econ Theory, 2000. **15**: p. 1-53.

43. Chandra A, K.P., Dorothy M, *Implications of Arrow's theorem in modeling of multiscale phenomena: an engineering approach.* NSF-DMMI grantee conference, Knoxville, TN, 2008.

44. Chandra A, R.S., *On removing Condorcet effects from pairwise election tallies.* Soc. Choice Welfare, 2013. **40**: p. 1143-1158.

45. Chandra A, K.O., Wu K-C, Hall M, Gillette J, *Prognosis of anterior cruciate ligament reconstruction: a data-driven approach.* Proc. R. Soc. A, 2015. **A 471**(20140526).

46. Sadd, M.H., *Elasticity Theory, Applications, and Numerics.* Elsevier, 2014.